\documentclass[journal]{IEEEtran}

\usepackage{cite}
\usepackage{amsmath,amssymb,amsfonts}
\usepackage{graphicx}
\usepackage{textcomp}
\usepackage[dvipsnames]{xcolor}
\usepackage{booktabs} % For professional looking tables
\usepackage{multirow}
\usepackage{siunitx}
\usepackage{url}
\usepackage{hyperref} % Clickable links without boxes

\begin{document}
	
	\title{CNN-based IoT Device Identification: \\ A Comparative Study on Payload vs. Fingerprint}
	
\author{Kahraman~Kostas\\ kahramankostas@gmail.com}% <-this % stops a space

	\maketitle
	
	% --- Abstract ---
	\begin{abstract}
		The proliferation of the Internet of Things (IoT) has introduced a massive influx of devices into the market, bringing with them significant security vulnerabilities. In this diverse ecosystem, robust IoT device identification is a critical preventive measure for network security and vulnerability management. This study proposes a deep learning-based method to identify IoT devices using the Aalto dataset. We employ Convolutional Neural Networks (CNN) to classify devices by converting network packet payloads into pseudo-images. Furthermore, we compare the performance of this payload-based approach against a feature-based fingerprinting method. Our results indicate that while the fingerprint-based method is significantly faster (approximately 10x), the payload-based image classification achieves comparable accuracy, highlighting the trade-offs between computational efficiency and data granularity in IoT security.
	\end{abstract}
	
	
	% --- Section 1: Introduction ---
	\section{Introduction}
	\label{sec:introduction}
	The rapid expansion of the Internet of Things (IoT) has transformed modern network infrastructures, connecting billions of heterogeneous devices ranging from smart home appliances to industrial sensors. However, this exponential growth has expanded the attack surface, making networks increasingly susceptible to cyber threats. Identifying unauthorized or vulnerable devices within a network is a fundamental step in securing IoT environments.
	
	Traditional device identification methods often rely on MAC addresses or static headers, which can be easily spoofed or encrypted. Consequently, recent research has shifted towards traffic analysis using Machine Learning (ML) and Deep Learning (DL) techniques. Specifically, converting network traffic data into visual representations (pseudo-images) to leverage the powerful feature extraction capabilities of Convolutional Neural Networks (CNNs) has emerged as a promising direction.
	
	In this study, we present a CNN-based identification framework applied to the Aalto dataset \cite{aalto2017dataset,miettinen2017iot}. We investigate two distinct approaches:
	\begin{enumerate}
		\item \textbf{Payload-to-Image:} Converting raw packet payloads into 2D pseudo-images.
		\item \textbf{Fingerprint-based:} Using extracted feature sets from IoTDevIDv1 \cite{kostas2021iotdevid1}.
	\end{enumerate}
	By comparing these methodologies, we aim to provide insights into the trade-offs between classification accuracy and computational overhead in real-time IoT monitoring.
	
	% --- Section 2: Related Work ---
	\section{Related Work}
	\label{sec:related_work}
	
	The application of CNNs in network security typically involves transforming network packet information into pseudo-images. Various encoding schemes have been proposed in the literature to map sequential network traffic into 2D structures suitable for image classification models.
	
\begin{figure}[htbp]
		\centering
		\includegraphics[width=0.9\linewidth]{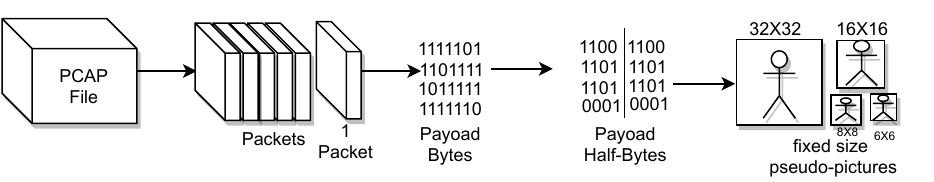}
		\caption{Overview of the image generation algorithm proposed by Lim et al. \cite{lim2019packet}.}
		\label{fig:lim_algo}
	\end{figure}

\begin{figure}[htbp]
		\centering
		\includegraphics[width=0.9\linewidth]{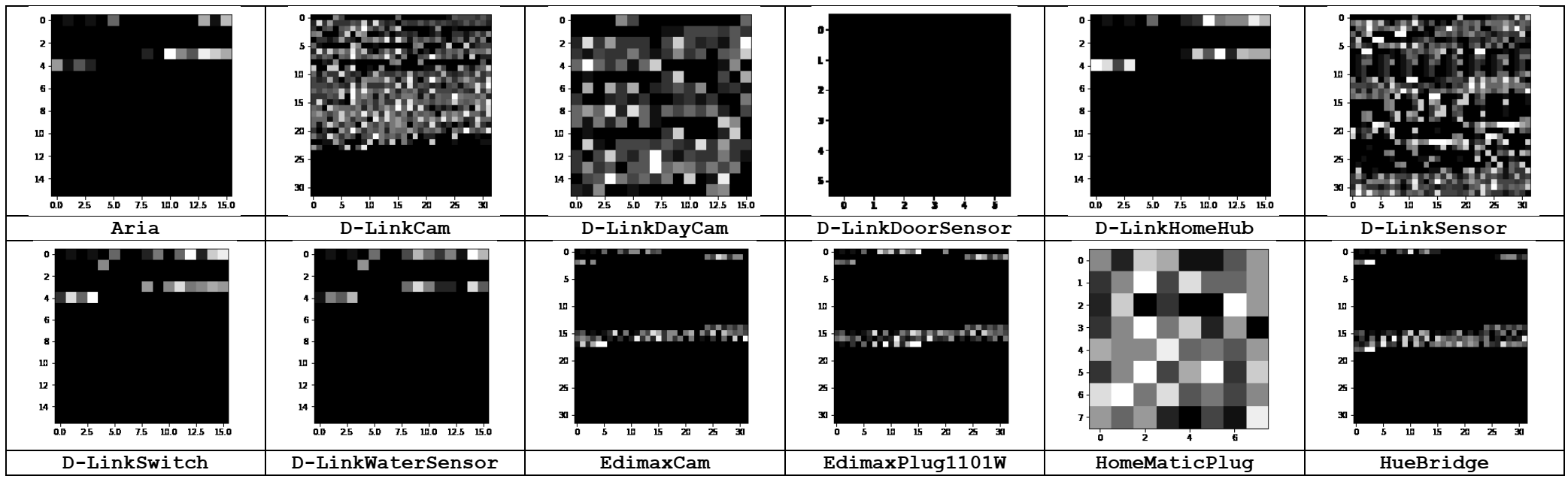}
		\caption{Sample pseudo-images generated from the first packet of 10 different devices based on \cite{lim2019packet}.}
		\label{fig:lim_samples}
	\end{figure}

	Lim et al. \cite{lim2019packet} proposed a method to convert network data into pseudo-images by processing the payload portion of the packets. In their approach, the payload is first converted into binary, then grouped into 4-bit nibbles. These nibbles are subsequently transformed into decimal values, which serve as pixel intensities. Depending on the payload size, pseudo-images of varying dimensions (e.g., $6 \times 6$, $8 \times 8$, $16 \times 16$, or $32 \times 32$) are generated. Zero-padding is applied to match the nearest standard image size, while payloads exceeding the maximum threshold are truncated. A conceptual representation of this algorithm is shown in Fig. \ref{fig:lim_algo}, and sample generated images are presented in Fig. \ref{fig:lim_samples}.
	
\begin{figure}[htbp]
		\centering
		\includegraphics[width=0.9\linewidth]{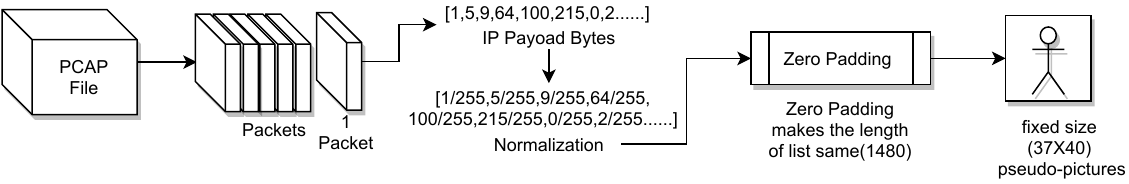}
		\caption{Preprocessing and image generation workflow by Lotfollahi et al. \cite{lotfollahi2020deep}.}
		\label{fig:lotfollahi_algo}
	\end{figure}

	Lotfollahi et al. \cite{lotfollahi2020deep} utilized PCAP files for both CNN and Stacked Autoencoders (SAE). Their preprocessing pipeline involves removing the Ethernet header and standard control packets (e.g., DNS, three-way handshake) that do not carry distinctive payload information. To ensure uniform input size, UDP packets and IP payloads smaller than 1480 bytes are zero-padded. Normalization is then performed by dividing each byte value by 255, resulting in a pseudo-image of 1480 pixels ($37 \times 40$). The process is illustrated in Fig. \ref{fig:lotfollahi_algo}.
	
	\begin{figure}[htbp]
		\centering
		\includegraphics[width=\linewidth]{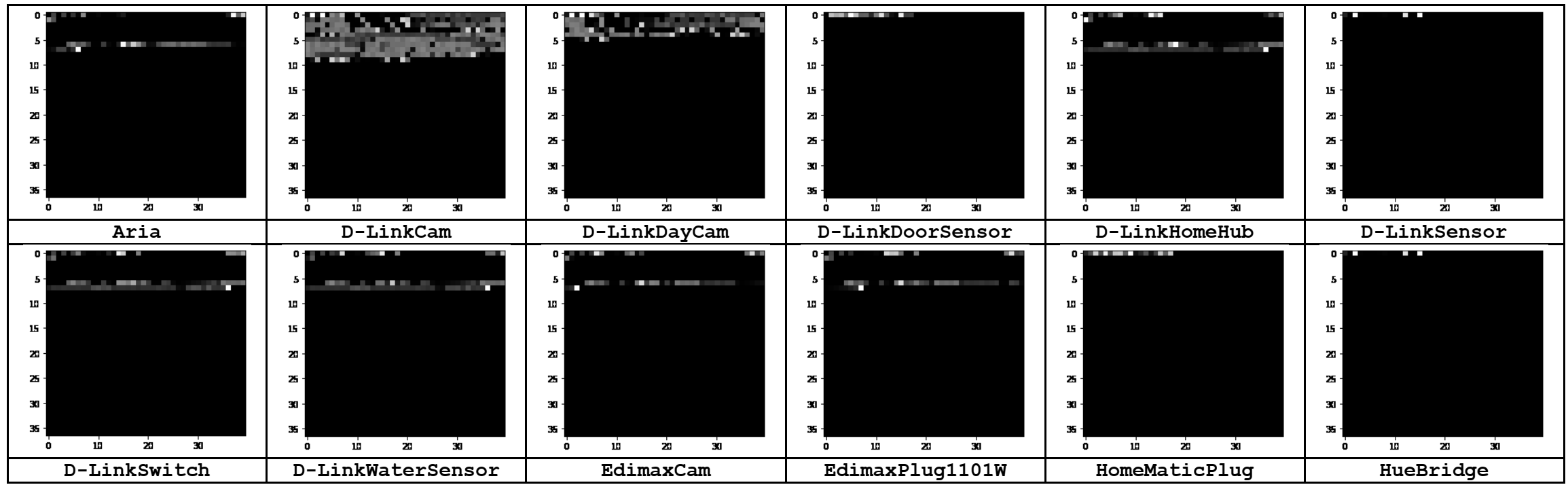}
		\caption{Visual representations of the first packet for 10 different devices based on the study of Lotfollahi et al. \cite{lotfollahi2020deep}.}
		\label{fig:lotfollahi_samples}
	\end{figure}
	
\begin{figure}[htbp]
		\centering
		\includegraphics[width=0.9\linewidth]{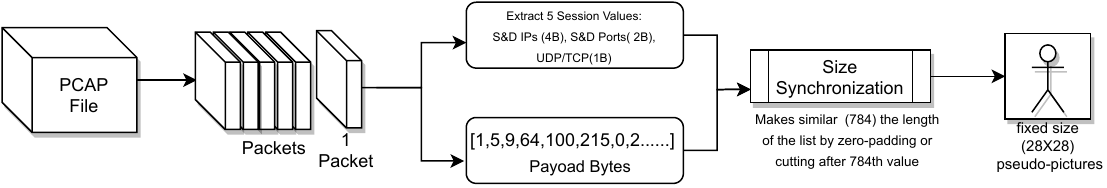}
		\caption{Data processing flow proposed by Wang et al. \cite{wang2017end}.}
		\label{fig:wang_algo}
	\end{figure}

	Wang et al. \cite{wang2017end} explored four different data representation groups: Session+All, Session+L7, Flow+All, and Flow+L7. They converted these data groups into $28 \times 28$ (784 pixels) pseudo-images. In their definition, a "Session" refers to bidirectional 5-tuple flows, whereas "Flow" is unidirectional. "L7" denotes the application layer payload. Fig. \ref{fig:wang_algo} depicts their conversion process.
	
\begin{figure}[htbp]
		\centering
		\includegraphics[width=\linewidth]{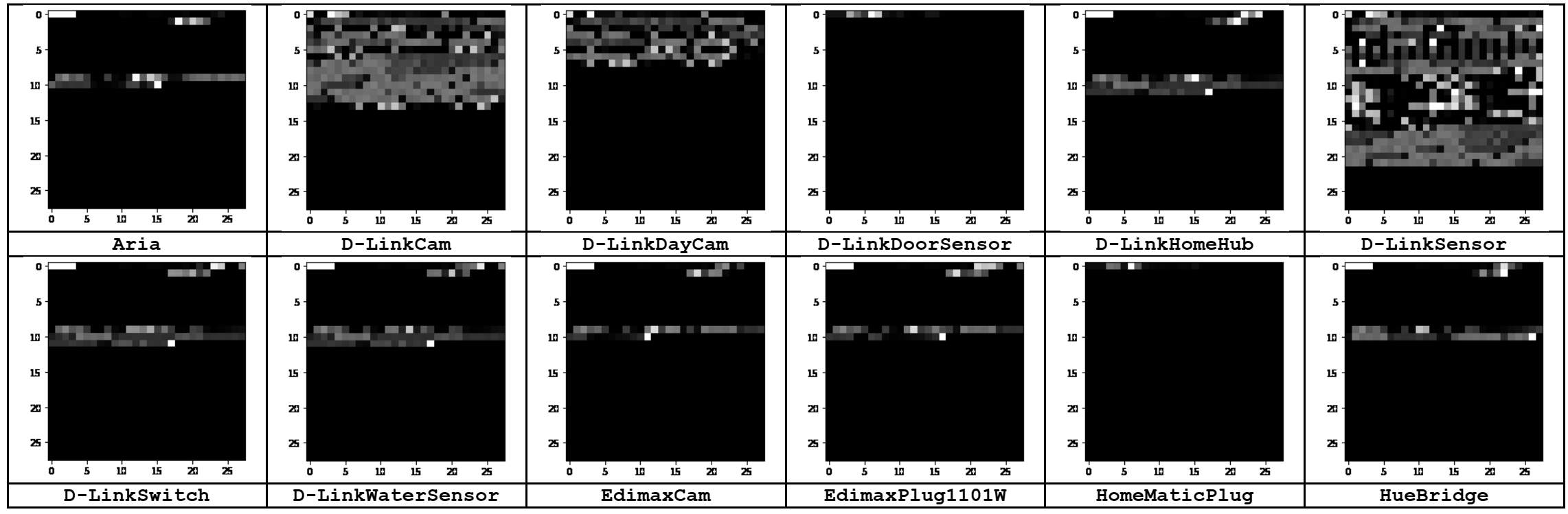}
		\caption{Pseudo-images representing the first packet for 10 distinct devices, generated using the method in \cite{wang2017end}.}
		\label{fig:wang_samples}
	\end{figure}

	In our study, we adopted the \textit{Session+L7} approach from Wang et al., as it focuses on the application layer payload which is rich in device-specific signatures. The first 13 bytes of our generated images represent the session headers (IPs, Ports, Protocol), while the remaining 771 bytes represent the payload (see Fig. \ref{fig:wang_samples}). Unlike Wang et al., we prioritize payload data over header features to evaluate the feasibility of deep packet inspection (DPI) via CNNs.
	
	% --- Section 3: Methodology and Results ---
	\section{Proposed Method and Evaluation}
	\label{sec:methodology}
	
\begin{figure}[htbp]
		\centering
		\includegraphics[width=0.8\linewidth]{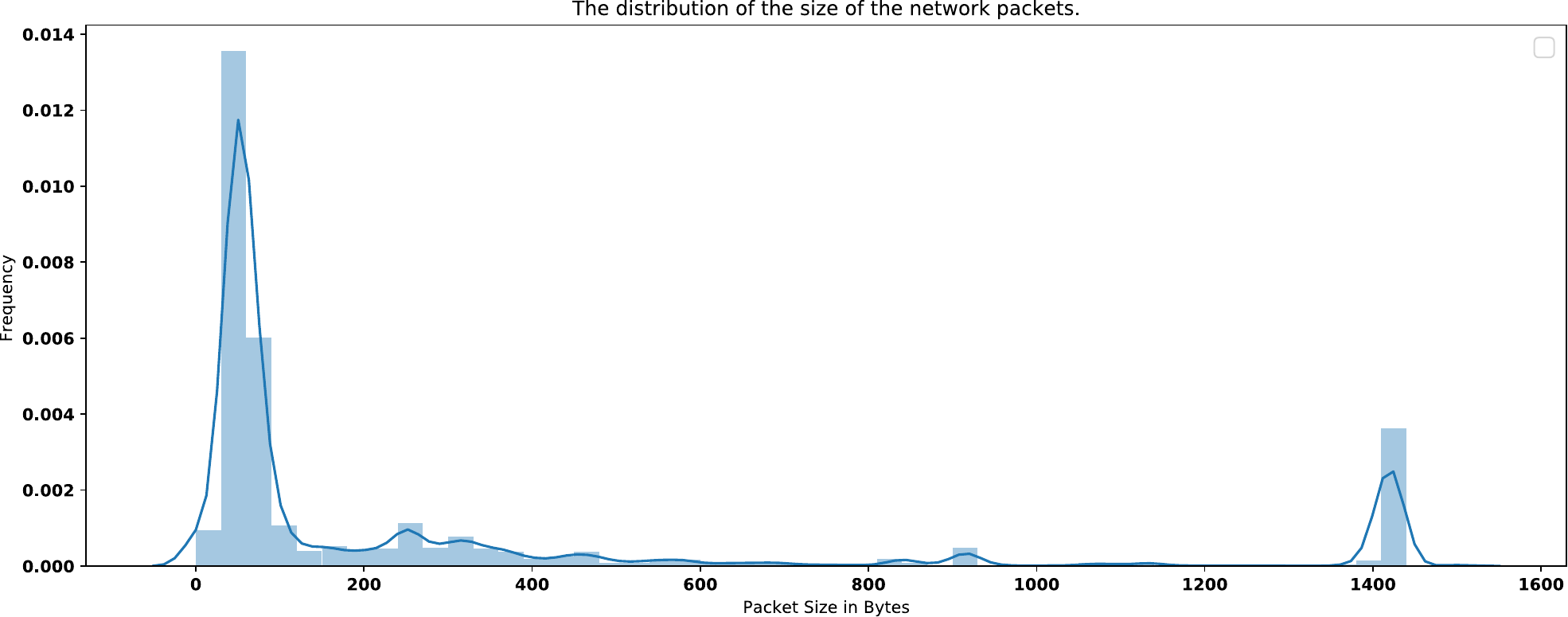}
		\caption{Distribution of network packet sizes in the dataset.}
		\label{fig:sizes}
	\end{figure}

	Our approach aims to standardize the data input for a single CNN architecture, avoiding the multi-model complexity seen in Lim et al. \cite{lim2019packet}. We focus on the first 784 bytes of the packet, as our analysis reveals that the majority of IoT network packets fall within the 0 to 800-byte range (see Fig. \ref{fig:sizes}).
	
	\subsection{Data Preprocessing}
	We evaluate two distinct feature extraction strategies:
	\begin{enumerate}
		\item \textbf{Payload-based CNN:} All packet headers are stripped. Payloads are converted into a $28 \times 28$ matrix. Empty packets (e.g., SYN/ACK without data) are discarded. Payloads smaller than 784 bytes are zero-padded, while larger ones are truncated.
		\item \textbf{Fingerprint-based CNN:} Pseudo-images are generated from the feature sets defined in IoTDevIDv1 \cite{kostas2021iotdevid1}. The generation script is available publicly\footnote{Available at: \url{https://github.com/kahramankostas/CNN-based-IoT-Device-Identification}}.
	\end{enumerate}
	
	\subsection{Model Configuration}
	To determine the optimal hyperparameters, we utilized \textit{AutoKeras}~\cite{jin2019auto}, an AutoML library. The architectural details of the CNN models employed for both approaches are summarized in Table~\ref{tab:model_comparison}

	\begin{table}[htbp]
		\centering
		\caption{Architectural Details and Parameter Comparison of the Proposed Models}
		\label{tab:model_comparison}
		\begin{tabular}{@{}lcc@{}}
			\toprule
			\textbf{Layer (Type)} & \textbf{\begin{tabular}[c]{@{}c@{}}Fingerprint Model\\ (Output Shape)\end{tabular}} & \textbf{\begin{tabular}[c]{@{}c@{}}Payload Model\\ (Output Shape)\end{tabular}} \\ \midrule
			Input & $(5, 5, 1)$ & $(28, 28, 1)$ \\
			Conv2D (1) & $(5, 5, 32)$ & $(26, 26, 32)$ \\
			Conv2D (2) & $(5, 5, 32)$ & $(24, 24, 64)$ \\
			MaxPooling2D & $(2, 2, 32)$ & $(12, 12, 64)$ \\
			Conv2D (3) & $(2, 2, 32)$ & - \\
			Conv2D (4) & $(2, 2, 32)$ & - \\
			Dropout & - & $(12, 12, 64)$ \\
			Flatten/Pooling & Global Avg. Pool & Flatten (9216) \\
			Dense (Output) & $(27)$ & $(27)$ \\ \midrule
			\textbf{Total Params} & \textbf{28,955} & \textbf{267,678} \\ \bottomrule
		\end{tabular}
	\end{table}
	\subsection{Results}
\begin{figure}[htbp]
		\centering
		\includegraphics[width=0.9\linewidth]{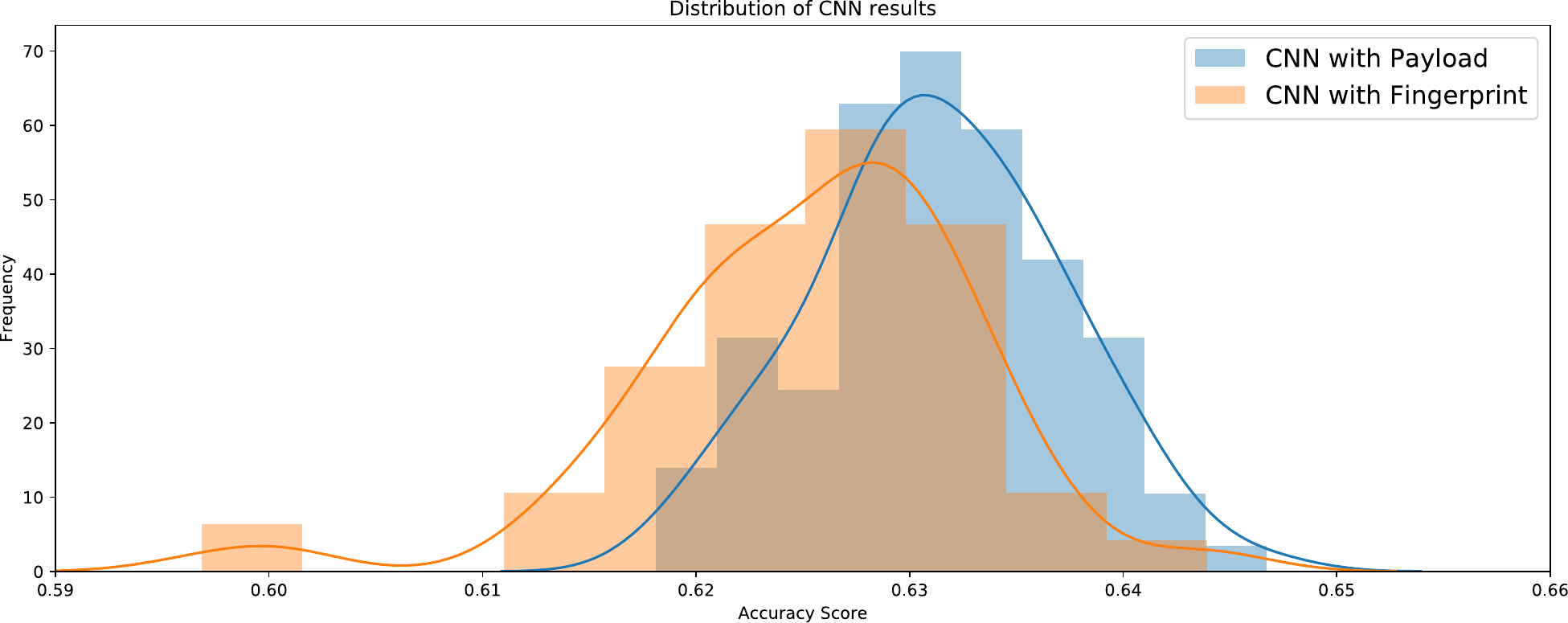}
		\caption{Accuracy distribution comparison between Payload and Fingerprint approaches.}
		\label{fig:CNNResults}
	\end{figure}

	The performance comparison between the two approaches is summarized in Table \ref{table:CNNResults}. The distribution of accuracy scores across multiple runs is visualized in Fig. \ref{fig:CNNResults}.
	
	\begin{table}[htbp]
		\caption{Performance Comparison: Payload vs. Fingerprint}
		\label{table:CNNResults}
		\centering
		\begin{tabular}{lcc}
			\toprule
			\textbf{Metric} & \textbf{CNN with Payload} & \textbf{CNN with Fingerprint} \\
			\midrule
			Accuracy & 0.631 & 0.625 \\
			Training Time (s) & 209.93 & 20.81 \\
			\bottomrule
		\end{tabular}
		\vspace{1ex}
		\\ \footnotesize{Note: Significance level is 0.05 for the Mann-Whitney U test.}
	\end{table}
	
As presented in Table \ref{table:CNNResults}, the payload-based method achieves a marginally higher accuracy (0.631) compared to the fingerprint-based method (0.625). However, a critical trade-off is observed regarding computational efficiency. While the Fingerprint-based method shows a negligible accuracy drop of only 0.6\%, it reduces the training time by approximately 90\% (from 209.9s to 20.8s). Furthermore, as detailed in Table \ref{tab:model_comparison}, the Fingerprint model requires significantly fewer trainable parameters (28,955) compared to the Payload model (267,678). This substantial reduction in model complexity suggests that while raw payload analysis offers slightly better granularity, the feature-based fingerprinting approach provides a far more efficient solution for resource-constrained IoT edge devices where memory and processing power are limited
	
	% --- Section 4: Conclusion ---
	\section{Conclusion}
	\label{sec:conclusion}
	
	In this study, we addressed the challenge of IoT device identification by evaluating a deep learning framework based on Convolutional Neural Networks (CNN). We conducted a comparative analysis of two distinct data representation strategies using the Aalto dataset: a raw payload-to-image conversion method and a feature-based fingerprinting approach.
	
	Our experimental results demonstrate that while the payload-based method achieves a marginally higher accuracy (63.1\%), the fingerprint-based method (62.5\%) offers a compelling alternative with significantly lower computational overhead. Specifically, the fingerprint approach reduced the training time by approximately 90\% (10x faster) with a negligible loss in accuracy. This finding is critical for resource-constrained IoT environments where real-time monitoring and rapid response are prioritized over marginal gains in classification precision.
	
	The significant reduction in computational overhead suggests that the Fingerprint-based approach is a viable candidate for real-time IoT device identification systems  deployed directly on IoT gateways, where deep packet inspection (payload analysis) might induce unacceptable latency.
	
	Future work will focus on validating these findings across larger and more diverse datasets, such as 		\href{https://iotanalytics.unsw.edu.au/iottraces}{UNSW-DI}\cite{sivanathan2018classifying} 
	\href{https://iotanalytics.unsw.edu.au/attack-data}{UNSW-AD}\cite{hamza2019detecting} 
	\href{https://github.com/netlab-stevens/LSIF}{LSIF}\cite{charyyev2020iot} 
	\href{https://moniotrlab.khoury.northeastern.edu/publications/imc19/}{MonIoTr}\cite{ren-imc19} 
	\href{https://www.unb.ca/cic/datasets/iotdataset-2022.html}{CIC-IoT-22}\cite{CIC} 
	. Additionally, we plan to explore hybrid architectures that can dynamically switch between lightweight fingerprinting for general monitoring and deep payload inspection for ambiguous cases.
	
	% --- References ---
	\bibliographystyle{IEEEtran}
	\bibliography{simple}

\end{document}